\documentclass[%
 reprint,
 superscriptaddress, 
 amsmath,amssymb,
 aps,
 prl, 
]{revtex4-2}

\usepackage{graphicx}
\usepackage{dcolumn}
\usepackage{bm}
\usepackage{xcolor}
\usepackage[colorlinks=true,linkcolor=blue,citecolor=blue,urlcolor=blue]{hyperref}

\begin{document}

\title{A Unified Wake Topology Map for He II Counterflow Past a Cylinder}

\author{Yingxuan Hu}
\author{Wenling Huang}
\author{Shihao Yang}
\affiliation{Institute of Refrigeration and Cryogenics, Zhejiang University, Hangzhou 310027, China}

\author{Limin Qiu}
\email[Email: ]{limin.qiu@zju.edu.cn}
\affiliation{Institute of Refrigeration and Cryogenics, Zhejiang University, Hangzhou 310027, China}
\affiliation{Zhejiang Key Laboratory of Refrigeration and Cryogenic Technology, Hangzhou 310027, China}

\author{Wei Guo}
\email[Email: ]{wguo@magnet.fsu.edu}
\affiliation{National High Magnetic Field Laboratory, 1800 East Paul Dirac Drive, Tallahassee, Florida 32310, USA}
\affiliation{Mechanical and Aerospace Engineering Department, FAMU-FSU College of Engineering, Florida State University, Tallahassee, Florida 32310, USA}

\author{Shiran Bao}
\email[Email: ]{srbao@zju.edu.cn}
\affiliation{Institute of Refrigeration and Cryogenics, Zhejiang University, Hangzhou 310027, China}
\affiliation{Zhejiang Key Laboratory of Refrigeration and Cryogenic Technology, Hangzhou 310027, China}

\date{\today}

\begin{abstract}
Thermal counterflow of superfluid $^4$He past a cylinder produces quasi-steady eddies not only downstream but also anomalously upstream. However, the mechanism and organizing principles behind the observed multistable wake topologies (0-, 2-, 4-, and 6-vortex states) have remained unclear. We show that the full spectrum of reported normal-fluid wake states is captured numerically with a two-fluid model coupled to Vinen's vortex-line-density equation. Our simulations further reveal that the superfluid component can also develop anomalous upstream eddies, a feature not previously reported. We trace these behaviors to a self-organized zone of enhanced mutual-friction dissipation near the cylinder shoulders that reshapes the effective obstacle, drives upstream eddies in both components, and suppresses intrinsic wake oscillations in the normal fluid. Guided by this mechanism, we perform systematic parameter scans and construct a unified phase diagram in terms of the normal-fluid Reynolds number $Re_n$ and a dimensionless interaction number $N$, separating inertia- and mutual-friction-controlled transitions and delineating the parameter windows for the discrete wake topologies. These results turn a striking phenomenology into a predictive map and establish mutual-friction feedback as a robust route to unusual wake structures in quantum fluids.
\end{abstract}

\maketitle

Flow around a circular cylinder is among the most paradigmatic problems in fluid mechanics. In classical viscous fluids, the canonical wake sequence—from steady recirculation to the formation of the K\'arm\'an vortex street—has been mapped out in detail and organized by well-defined control parameters such as the Reynolds number ($Re$)~\cite{Batchelor2000,Williamson1996}. Replacing the working fluid with superfluid $^4$He (He~II) below the lambda point at 2.17~K fundamentally transforms the flow physics. He~II is a quantum fluid described by Landau–Tisza two-fluid model~\cite{Landau1987}: an inviscid superfluid component coexists with a viscous normal component composed of thermal quasiparticles. The superfluid supports quantized vortex lines with circulation fixed to $\kappa\simeq 9.97\times10^{-4}$~cm$^2$/s~\cite{Donnelly1991}. As vortices move relative to the normal fluid, mutual friction between the two fluids arises from scattering of thermal quasiparticles off the vortex cores~\cite{Vinen1957a,Vinen1957b,tang2023imaging}. This coupling allows the evolving vortex tangle to feed back on the two-fluid motion and reshape wake formation, opening the door to flow states with no classical analogue.

A particularly striking manifestation occurs in thermally driven counterflow. When a heat flux is applied, the normal fluid carries entropy away from the heater while the superfluid flows oppositely to maintain zero net mass flux~\cite{VanSciver2012}. Zhang \textit{et al.} discovered that, in such counterflow past a cylinder, large-scale quasi-steady vortex pairs can appear not only downstream but also anomalously upstream of the obstacle~\cite{Zhang2005}. This behavior sharply contrasts with the classical cylinder wake, where steady eddies reside only downstream. Subsequent experiments by different groups extended these observations across a range of conditions, revealing that the wake is not described by a single configuration but instead exhibits pronounced multistability: depending on parameters, the flow can settle into states with no vortices, or with two, four, or six large-scale vortices~\cite{VanSciver2007,Chagovets2013,Duda2014,Fuzier2006}. These findings suggest that mutual-friction coupling can reorganize the wake topology and introduces robust new states.


Considerable effort has been devoted to explaining the anomalous upstream eddies. Sergeev \textit{et al.} introduced an ideal point-vortex model showing that stationary upstream vortex configurations can exist for specific circulation values~\cite{Sergeev2009}. While insightful, this inviscid framework omits the mutual-friction feedback that are central to He~II counterflow. Later, Soulaine \textit{et al.} achieved the first numerical reproduction of the anomalous four-vortex wake using two-fluid simulations and showed that its onset is governed by a dimensionless mutual-friction intensity~\cite{Soulaine2017}. Later work examined how heat flux, boundary conditions, and modeling choices influence the evolution and stability of this four-vortex state~\cite{Li2023,Yousefi2025}. Despite these advances, a central gap remains: experiments demonstrate a family of distinct, long-lived wake topologies and transitions among them, whereas most numerical studies have focused on a single representative state. A unified framework is still lacking that (i) captures all the observed wake topologies, and (ii) organizes the transitions into a predictive phase diagram. Meanwhile, the superfluid wake has remained largely unexplored: existing reports focus on normal-fluid streamline patterns, leaving open whether the superfluid can also develop analogous wake structures. 

\begin{figure*}[t]
\includegraphics[width=\textwidth]{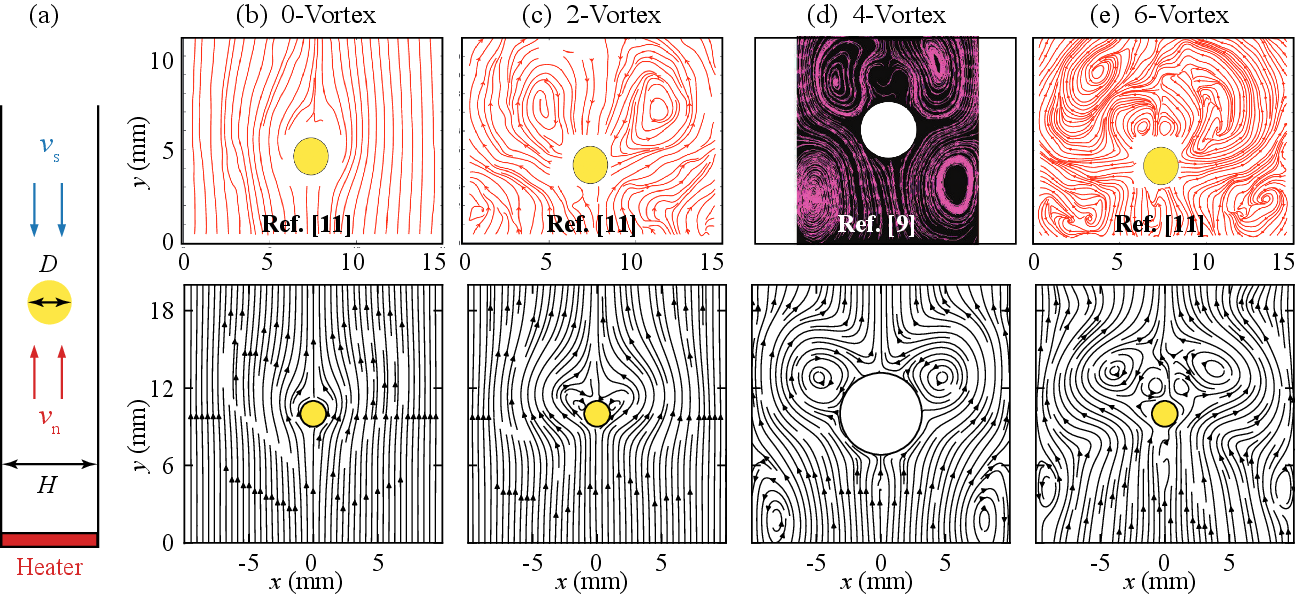}
\caption{\label{fig:domain} (a) Schematic of the computational domain. (b)--(e) Comparison between experimentally reconstructed streamline patterns from particle tracks (top row)~\cite{Zhang2005,Chagovets2013} and simulated normal-fluid streamlines (bottom row) at matched $T$, $q$, and blockage ratio $B\equiv D/H$. Parameters are (b) $T=1.94$~K, $q=50$~mW/cm$^2$, $B=10\%$; (c) $T=1.94$~K, $q=170$~mW/cm$^2$, $B=10\%$; (d) $T=2.03$~K, $q=1120$~mW/cm$^2$, $B=31.75\%$; (e) $T=2.10$~K, $q=167$~mW/cm$^2$, $B=10\%$.}
\end{figure*}

In this Letter, we fill this gap using a unified continuum framework~\cite{Hu2025a,Hu2025b,bao2021transient}. Specifically, we solve the He~II two-fluid equations coupled to Vinen's vortex-line-density equation, so that the vortex tangle and the mutual friction it mediates evolve self-consistently with the flow. Within this description, we reproduce the experimentally observed multistable normal-fluid wake spectrum and further show that the superfluid component can also develop anomalous upstream eddies, a feature not previously reported. We trace this coupled behavior to a self-organized, strongly dissipative region near the cylinder shoulders that, through nonlinear mutual-friction feedback, reshapes the effective obstacle and drives the upstream recirculation that sustains the anomalous eddies. Building on this mechanism, we construct a unified phase diagram in terms of selected dimensionless control parameters, which organizes the discrete wake topologies and separates inertia- and mutual-friction-controlled transitions, providing a mechanism-based framework for wake-topology bifurcations in He~II counterflow.

As illustrated in Fig.~\ref{fig:domain}(a), the computational domain is a 2D channel of length $L_c$ and width $H$ containing a circular cylinder of diameter $D$. We model the He II flow using the two-fluid hydrodynamics~\cite{Landau1987, Nemirovskii1983}, in which the normal and superfluid components with velocities $\mathbf{v}_n$ and $\mathbf{v}_s$ (densities $\rho_n$ and $\rho_s$) are coupled through a volumetric mutual-friction force $\mathbf{F}_{ns}$ as~\cite{Vinen1957b,Nemirovskii2020,Tsubota2017}:
\begin{align}
\rho_n\frac{D_n \mathbf{v}_n}{Dt}= -\frac{\rho_n}{\rho}\nabla p-\rho_s\nabla\mu+\nabla\cdot\left(\eta_n\nabla\mathbf{v}_n\right)-\mathbf{F}_{ns},\label{eq:momentum_n}\\
\rho_s\frac{D_s \mathbf{v}_s}{Dt}= -\frac{\rho_s}{\rho}\nabla p+\rho_s\nabla\mu+\mathbf{F}_{ns},\label{eq:momentum_s}\\
\frac{\partial}{\partial t}\left(\rho s\right)+\nabla\cdot\left(\rho s\,\mathbf{v}_n\right)= \frac{1}{T}\mathbf{F}_{ns}\cdot\mathbf{v}_{ns},\label{eq:entropy}
\end{align}
where $D_n/Dt\equiv \partial/\partial t+\mathbf{v}_n\cdot\nabla$ and $D_s/Dt\equiv \partial/\partial t+\mathbf{v}_s\cdot\nabla$, $\rho=\rho_n+\rho_s$ is the total density, $p$ is the pressure, $\eta_n$ is the normal-fluid viscosity, $s$ is the specific entropy, and $\mathbf{v}_{ns}=\mathbf{v}_n-\mathbf{v}_s$ is the relative velocity. The chemical potential gradient is given by $\nabla\mu=-s\nabla T-\frac{\rho_n}{2\rho}\nabla\left|\mathbf{v}_{ns}\right|^2$~\cite{Landau1987}.

At the coarse-grained level, the mutual-friction force is modeled as $\mathbf{F}_{ns}=\frac{\rho_n\rho_s}{3\rho}\,B_L(T)\,L\,\mathbf{v}_{ns}$, where $B_L(T)$ is the tabulated mutual-friction coefficient~\cite{Donnelly1998} and $L(\mathbf{r},t)$ is the vortex-line density, whose evolution is described by Vinen's equation~\cite{Vinen1957a,Vinen1957b}:
\begin{equation}
\frac{\partial L}{\partial t} + \nabla \cdot (\mathbf{v}_L L)
= \alpha_v |\mathbf{v}_{ns}|\, L^{3/2} - \beta_v L^2 + \gamma_v |\mathbf{v}_{ns}|^{5/2},
\label{eq:vinen}
\end{equation}
where $\alpha_v$, $\beta_v$, and $\gamma_v$ are temperature-dependent empirical coefficients, and the vortex drift velocity is taken as $\mathbf{v}_L \simeq \mathbf{v}_s$~\cite{Vinen1957a,Vinen1957b,Wang1987,kondaurova2014structure}. Note that instead of solving Eq.~(\ref{eq:vinen}), Soulaine \emph{et al.} assumed instantaneous local equilibrium value $L=(\alpha_v/\beta_v)^2|\mathbf{v}_{ns}|^2$ in their work~\cite{Soulaine2017}. This reduction is valid only when the vortex relaxation dynamics is fast compared with flow advection~\cite{Nemirovskii2013}.

\begin{figure*}[t]
\includegraphics[width=\textwidth]{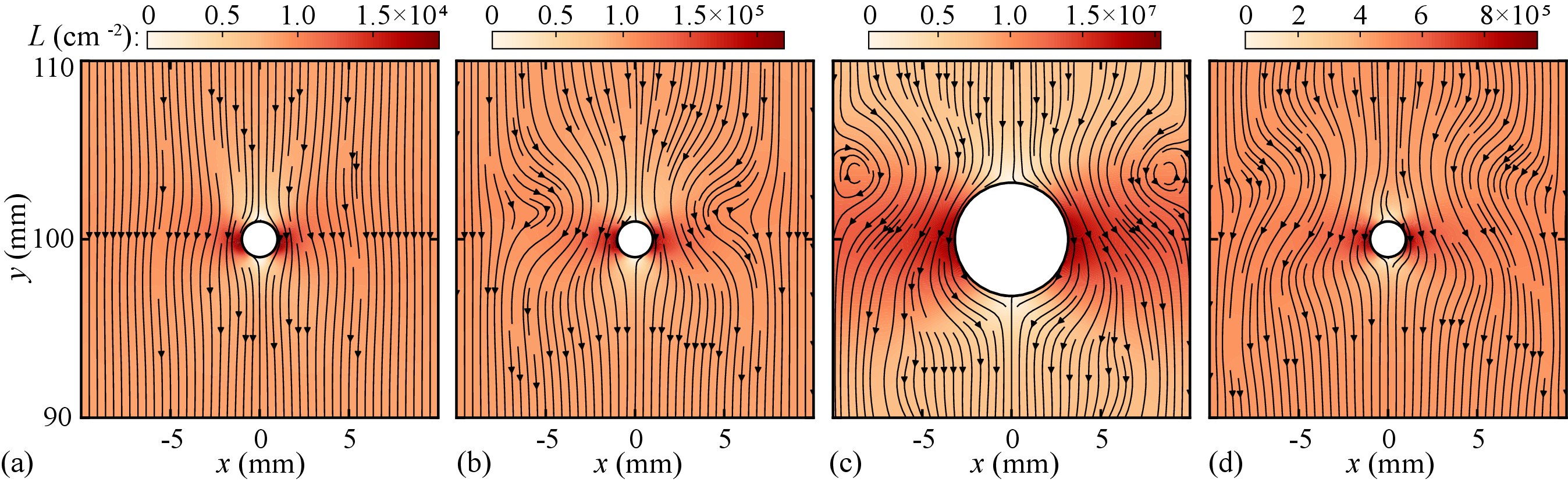}
\caption{\label{fig:mechanism} Simulated superfluid streamlines overlaid on a colormap of the vortex-line density $L(x,y)$. Panels (a)-(d) use the same parameters $T$, $q$, and $B$ as Fig.~\ref{fig:domain}(b)-(e), respectively. The eddies in (c) are upstream relative to the superfluid flow direction.}
\end{figure*}

We impose counterflow by prescribing the inlet normal-fluid velocity using the applied heat flux $q$ as $|\mathbf{v}_{n,0}|=q/(\rho s T)$, together with $\mathbf{v}_{s,0}=-(\rho_n/\rho_s)\mathbf{v}_{n,0}$ to enforce zero net mass flux. On the cylinder surface and channel walls, we apply no slip for the normal fluid, $\mathbf{v}_n=\mathbf{0}$, and impermeability for the superfluid, $\mathbf{v}_s\cdot\hat{\mathbf{n}}=0$. Although this continuum model does not resolve individual quantized vortices~\cite{barenghi2006depolarization,baggaley2012sensitivity,yui2022universal}, it has been extensively validated as a reliable description of the coupled two-fluid dynamics on length scales larger than the mean intervortex spacing~\cite{Nemirovskii2020,sanavandi2022boiling,inui2023boiling,Tsubota2017,Galantucci2015}. Numerical details (discretization, boundary conditions, and convergence tests) are provided in the Supplemental Material (SM)~\cite{SM}.


To validate the model and demonstrate its predictive capability, we performed numerical simulations at the same bath temperature $T$, heat flux $q$, and channel blockage ratio $B\equiv D/H$ as in the experiments of Refs.~\cite{Zhang2005,Chagovets2013}. Fig.~\ref{fig:domain}(b)--(e) shows a direct comparison between experimentally reconstructed streamline patterns from tracer tracks with the simulated normal-fluid streamline topologies under these matched conditions. The simulations reproduce the full experimentally observed spectrum of quasi-steady wake states, namely the 0-, 2-, 4-, and 6-vortex configurations. At low $q$, the wake is vortex-free and nearly potential. As $q$ increases, a steady downstream recirculation forms (2-vortex state). With further increase in $q$, the wake does not develop periodic vortex shedding as in classical cylinder flows; instead, the anomalous 4-vortex state emerges, marked by the formation of an additional upstream vortex pair. For small $B$, the wake can further organize into a symmetric 6-vortex configuration at sufficiently large $q$.

The fact that these discrete, multistable states emerge in the normal fluid within one continuum framework highlights vortex-mediated mutual friction as the key nonlinear mechanism controlling wake-topology selection. The essential ingredient is the strong spatial inhomogeneity of the counterflow around the cylinder. As the incoming flow is diverted, the counterflow is funneled through the two shoulder corridors, producing a locally large relative velocity $\mathbf{v}_{ns}$; because the vortex-line density rises rapidly with $\mathbf{v}_{ns}$ toward its local-equilibrium scaling $L_{\mathrm{eq}}\propto|\mathbf{v}_{ns}|^2$~\cite{Vinen1957b,SM}, the vortex tangle concentrates near the shoulders. This concentration is directly visible in Fig.~\ref{fig:mechanism}, which plots $L(x,y)$ together with the superfluid streamlines for the same $(T,q,B)$ as the corresponding normal-fluid cases in Fig.~\ref{fig:domain}(b)-(e). Since the mutual-friction force scales as $\mathbf{F}_{ns}\propto L \mathbf{v}_{ns}$, the shoulder-localized tangle creates a compact zone of enhanced dissipation that acts as a permeability barrier, effectively “thickening” the obstacle and redirecting the incoming normal flow. Together with the lateral confinement imposed by the channel walls, this diverted flow can close into an upstream recirculating cell and thereby sustains the anomalous upstream vortex pair. This phenomenon is a close analogue of elastic blockage in viscoelastic wakes, where a localized buildup of nonlinear stresses upstream of a confined obstacle similarly rejects the inflow and induces an upstream recirculation zone~\cite{Peng2023,Hopkins2022a,Hopkins2022b,Qin2019}.


The mutual-friction dissipation also explains the striking downstream wake stability. In a classical viscous cylinder wake, periodic K\'arm\'an shedding sets in already at low Reynolds number, $Re\simeq 47$, when inertia overcomes viscous diffusion~\cite{Williamson1996}. In sharp contrast, He~II counterflow experiments report a stable two-vortex configuration (Fig.~\ref{fig:domain}(c)) persisting up to $Re_n\equiv \rho_n v_{n,0}D/\eta_n\approx 1554$~\cite{Chagovets2013}. This dramatic stabilization arises because, as $q$ increases, the mutual friction $\mathbf{F}_{ns}$ rapidly overwhelms the normal-fluid viscous term $\nabla\cdot\left(\eta_n\nabla\mathbf{v}_n\right)$ across the computational domain (see SM~\cite{SM}). The separated shear layers therefore undergo strong mutual-friction damping, suppressing the shear-layer instability that drives K\'arm\'an vortex shedding~\cite{Bertolaccini2017}.

\begin{figure*}[t]
\includegraphics[width=\textwidth]{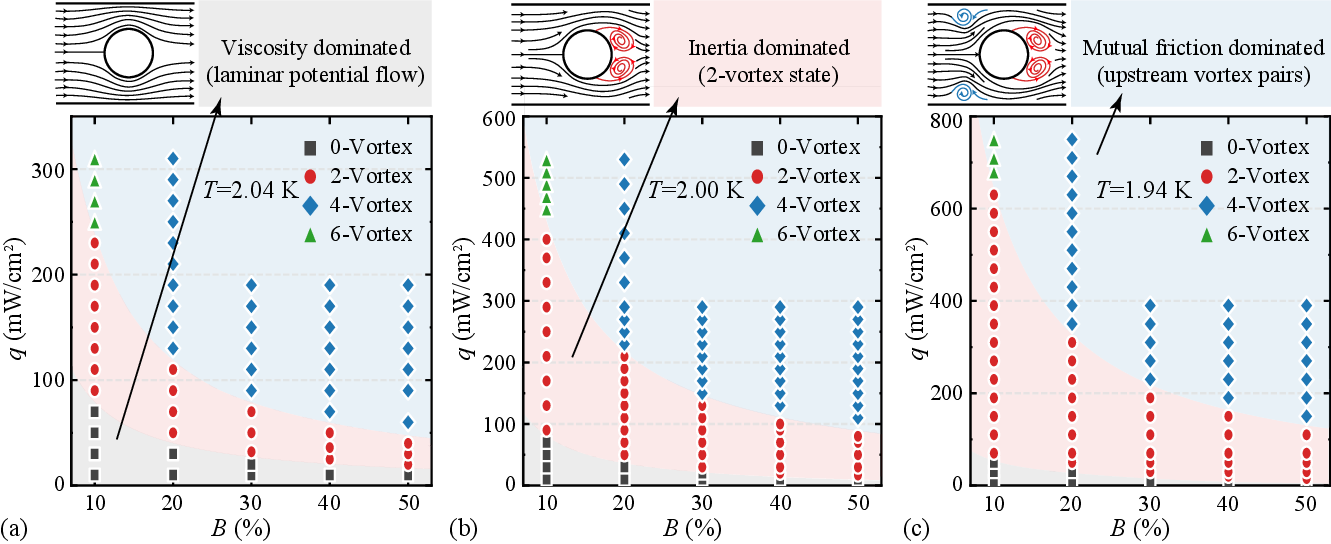}
\caption{\label{fig:phase_qB} Wake-topology phase diagrams in the $(q,B)$ plane obtained from systematic sweeps at (a) $T=2.04$ K, (b) $T=2.00$ K, and (c) $T=1.94$ K. Shaded backgrounds indicate the dominant physical mechanism, separating viscosity-dominated (gray), inertia-dominated (red), and mutual-friction-dominated (blue) regimes. Symbols mark the observed wake topologies.}
\end{figure*}

In Fig.~\ref{fig:mechanism}, we also show the superfluid response under the same conditions. Although the superfluid is inviscid, it can sustain eddy-like, coarse-grained circulation on length scales larger than the mean intervortex spacing, where partial vortex polarization and bundling generate a nonzero averaged vorticity~\cite{Babuin2014}. As $q$ increases, the resulting mutual-friction barrier increasingly deflects the superfluid streamlines and promotes an upstream recirculation tendency. However, because the superfluid does not satisfy a no-slip condition at solid boundaries, the near-wall superfluid speed remains large and readily erodes closed recirculating cells; robust upstream superfluid eddies therefore appear only under strong confinement (large $B$), where the wall geometry cooperates with the shoulder barrier to trap a recirculating cell (see Fig.~\ref{fig:mechanism}(c)). By contrast, the superfluid shows no detached downstream wake analogous to the normal-fluid recirculation, consistent with the absence of a viscous boundary layer on the cylinder and hence the lack of boundary-layer separation as a wake-forming mechanism~\cite{Batchelor2000}.

To locate transitions among the discrete wake states and place them on a common footing, we construct a wake-topology phase diagram by scanning the control-parameter space $(T,q,B)$. We surveyed the temperature range $T=1.9$--$2.1$~K and performed systematic sweeps in $q$ and $B$ to resolve the wake-state transitions. Fig.~\ref{fig:phase_qB} presents the resulting normal-fluid wake topology maps at three representative temperatures, $T=1.94$~K, 2.00~K, and 2.04~K. For each displayed $(T,B)$, we increased $q$ in steps of 20~mW/cm$^2$ and refined the increment to 5~mW/cm$^2$ near transition thresholds.

Across all three temperatures, Fig.~\ref{fig:phase_qB} exhibits the same organizing trends. At fixed blockage ratio $B$, increasing $q$ drives a systematic progression from a viscous-dominated, nearly potential-flow state (0-vortex) to an inertia-driven separated wake characterized by a downstream vortex pair (2-vortex), and then to a mutual-friction-controlled regime in which an additional upstream vortex pair nucleates, producing the anomalous four-vortex topology. The six-vortex state, with two upstream vortices accompanied by four quasi-stable vortices in the wake, appears only in a restricted corner of the parameter space at the smallest $B$ and sufficiently large $q$. Meanwhile, increasing $B$ shifts the onset of both the two-vortex and four-vortex states to lower heat flux in every panel, demonstrating that stronger confinement makes wake separation and upstream-pair nucleation accessible at smaller imposed $q$.

\begin{figure}[t]
\includegraphics[width=1\columnwidth]{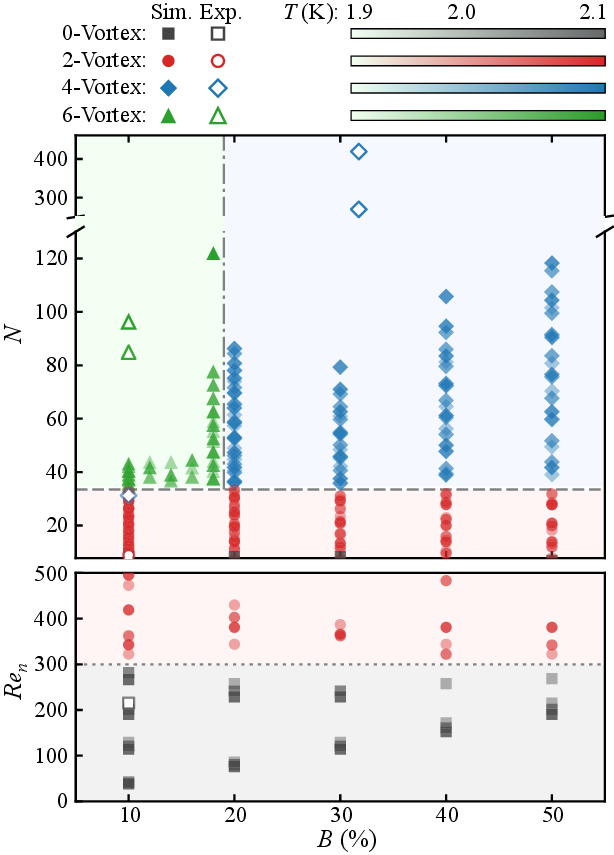}
\caption{\label{fig:phase_diagram} Stratified dimensionless phase diagram compiling all wake-topology points from Fig.~\ref{fig:phase_qB} together with the experimental data. Bottom: separation onset from the 0-vortex to 2-vortex state plotted in the $(Re_n,B)$ plane. Top: wake topology plotted as the interaction number $N$ versus $B$, showing the 2-vortex region below $N_c\simeq 33$, the 4-vortex region above $N_c$ at larger $B$, and the 6-vortex region above $N_c$ at small $B$. Open and filled symbols denote experimental and numerical results, respectively, and symbol opacity encodes the bath temperature $T$ (colorbar).
}
\end{figure}

The two sequential transitions in Fig.~\ref{fig:phase_qB} imply two distinct competitions. The first transition, from the 0-vortex state to the 2-vortex wake, is set by the balance between normal-fluid viscous diffusion and inertial advection, and is therefore naturally characterized by the normal-fluid Reynolds number $Re_n$. The second transition, from the 2-vortex wake to the anomalous 4-vortex topology, occurs once mutual friction becomes significant. To quantify this competition, we compare the characteristic force-density scales: $F_{in}\sim \rho_n v_{n,0}^2/D$ for inertial advection and $F_{ns}\sim\frac{\rho_n\rho_s}{3\rho}\kappa B_L L_0 v_{ns,0}$ for mutual friction. Using the Vinen local-equilibrium scaling $L_0\simeq (\alpha_v/\beta_v)^2 v_{ns,0}^2$ together with the inlet counterflow velocity $v_{ns,0}=(\rho/\rho_s)v_{n,0}$, the ratio $F_{ns}/F_{in}$ defines a dimensionless interaction number $N$ given by:
\begin{equation}
N=\left[\frac{\rho^2}{3\rho_s^2}B_L\kappa(\alpha_v/\beta_v)^2\right] v_{n,0}D.
\label{eq:interaction_number}
\end{equation}
This form mirrors Soulaine \textit{et al.}~\cite{Soulaine2017}: all temperature dependence is absorbed into a single prefactor, leaving the scaling controlled by the kinematic product $v_{n,0}D$.

Guided by this reasoning, we construct the stratified dimensionless phase diagram in Fig.~\ref{fig:phase_diagram} by replotting all normal-fluid wake-topology points, together with the experimental data, in two complementary projections. In the bottom panel, we show the onset of wake separation into the 2-vortex state in the $(Re_n,B)$ plane; the boundary collapses to a nearly constant threshold $Re_{n,c}\approx 300$, indicating a kinematic transition that is largely insensitive to $T$ and $B$ once viscosity and inertia are properly scaled. In the top panel, we show the onset of upstream-pair nucleation into the anomalous 4-vortex state in the $(N,B)$ plane; the boundary similarly collapses near a universal threshold $N_c\approx 33$, demonstrating that this bifurcation is governed by the balance between inertial advection and vortex-mediated mutual friction. The 6-vortex topology appears only for $N\gtrsim N_c$ and is further restricted to $B\lesssim 19\%$, indicating that, in addition to requiring sufficiently strong vortex feedback, this higher-order state is geometrically permitted only when confinement is weak enough to provide adequate downstream space for four quasi-stable wake vortices to coexist with the upstream pair~\cite{Sahin2004,Singha2010}.


Overall, these results turn the He~II counterflow cylinder wake from a long-standing visual curiosity into a quantitative benchmark for how mutual friction reshapes, stabilizes, and reorganizes macroscopic flow. The key advance is not merely reproducing multiple steady topologies, but demonstrating that their bifurcations can be predicted from a small set of measurable inputs within a continuum framework that retains the essential nonlinearity of vortex-mediated feedback. We note, however, an intrinsic limitation of coarse-grained two-fluid models of this class: quantized vortices enter only through a smoothed mutual-friction field, so effects associated with individual vortex lines are averaged out. In particular, a moving vortex can generate a macroscopic wake in the normal fluid~\cite{Kivotides-Science-2000, galantucci2020new, tang2023imaging, Galantucci-PRL-2026}, and in dense tangles such wakes may overlap and drive large-scale disturbances that could, in principle, promote normal-fluid turbulence even in the presence of strong mutual-friction damping~\cite{Mastracci-PRFluids-2019, Yui-PRL-2020}. These effects lie beyond the present model’s resolution. Nevertheless, within its range of validity the present study isolates the dominant mutual-friction-controlled mechanism and yields a quantitative wake-topology map, providing a practical baseline for interpreting measurements and suggesting strategies to tailor flow stability in He~II transport and thermal management where suppressing or triggering large-scale recirculation is desirable.

\begin{acknowledgments}
The authors thank S. Xiong for valuable discussions. S. Bao acknowledges support from the National Natural Science Foundation of China (Grant No. 52441601 and No. 52206028) and the Natural Science Foundation of Zhejiang Province (Grant No. LZ25E060002). L. Qiu acknowledges support from the Fundamental Research Funds for the Central Universities (Grant No. K20250227). W. Guo acknowledges support from the Gordon and Betty Moore Foundation through Grant DOI 10.37807/gbmf11567 and the National High Magnetic Field Laboratory at Florida State University, which is supported by the National Science Foundation Cooperative Agreement No. DMR-2128556 and the State of Florida.
\end{acknowledgments}

\bibliography{reference}

\end{document}


\title{Supplemental Material for: \\ A Unified Wake Topology Map for He II Counterflow Past a Cylinder}

\author{Yingxuan Hu}
\author{Wenling Huang}
\author{Shihao Yang}
\affiliation{Institute of Refrigeration and Cryogenics, Zhejiang University, Hangzhou 310027, China}

\author{Limin Qiu}
\email[Email: ]{limin.qiu@zju.edu.cn}
\affiliation{Institute of Refrigeration and Cryogenics, Zhejiang University, Hangzhou 310027, China}
\affiliation{Zhejiang Key Laboratory of Refrigeration and Cryogenic Technology, Hangzhou 310027, China}

\author{Wei Guo}
\email[Email: ]{wguo@magnet.fsu.edu}
\affiliation{National High Magnetic Field Laboratory, 1800 East Paul Dirac Drive, Tallahassee, Florida 32310, USA}
\affiliation{Mechanical and Aerospace Engineering Department, FAMU-FSU College of Engineering, Florida State University, Tallahassee, Florida 32310, USA}

\author{Shiran Bao}
\email[Email: ]{srbao@zju.edu.cn}
\affiliation{Institute of Refrigeration and Cryogenics, Zhejiang University, Hangzhou 310027, China}
\affiliation{Zhejiang Key Laboratory of Refrigeration and Cryogenic Technology, Hangzhou 310027, China}

\date{\today}

\maketitle

\section{Modeling Framework}
We simulate the He~II counterflow wake using the coupled two-fluid formulation described in the main text: the Landau--Tisza two-fluid equations are solved for the velocity fields of the normal and superfluid components, and the mutual-friction force is evaluated self-consistently from the local vortex-line density $L(\mathbf{r},t)$. The evolution of $L$ is governed by the Vinen equation (Eq.~(4) in the main text), which provides a coarse-grained description of vortex transport and of the local growth and decay of the vortex tangle under thermal counterflow driving.

In the Vinen equation, the term $\nabla\cdot(\mathbf{v}_L L)$ accounts for the drifting (advection) of vortices~\cite{Vinen1957a,Vinen1957b}. We take the vortex mean velocity to be the local superfluid velocity, $\mathbf{v}_L=\mathbf{v}_s$, as originally proposed by Vinen and widely adopted in subsequent two-fluid/Vinen implementations~\cite{Vinen1957a,Vinen1957b,Wang1987,kondaurova2014structure}. While alternative treatments of $\mathbf{v}_L$ (or, more generally, of the vortex-line-density flux) have been discussed~\cite{Nemirovskii2020}, we adopt this standard choice here because it provides a minimal and well-tested representation of vortex drift and captures the leading-order redistribution of $L$ by the computed superfluid advection. The source terms on the right-hand side of the Vinen equation describe the local evolution of the tangle: the first two terms account for vortex generation and decay, respectively, and the additional source term triggers the initial growth of $L$ under counterflow driving~\cite{Vinen1957b}. This form enables the simulation to capture the competition between advection, growth, and decay of the vortex tangle that underlies the mutual-friction feedback mechanism emphasized in the main text.

At solid boundaries, the normal fluid satisfies no-slip $\mathbf{v}_n=\mathbf{0}$ due to its viscosity. For the superfluid, because the effective tangential boundary condition depends on wall roughness and possible vortex pinning~\cite{Galantucci2015, Yui-2015-PRB,Stagg-2017-PRL}, which is not resolved in our coarse-grained model, we adopt the standard free-slip assumption and enforce only impermeability, $\mathbf{v}_s\cdot\hat{\mathbf{n}}=0$. Thermophysical properties of He~II (e.g., $\rho$, $s$, $\eta_n$, etc.) are obtained from the HePak database~\cite{hepak_arp}. The temperature-dependent coefficients in the Vinen equation ($\alpha_v$, $\beta_v$, $\gamma_v$) are taken from Vinen's compilations~\cite{Vinen1957b} and the recommendations of Kondaurova \textit{et al.}~\cite{Kondaurova2017}. All properties and empirical coefficients are updated dynamically using the local temperature field during the simulation.

\begin{table*}[t]
\caption{\label{tab:grid} Summary of the grid-independence test performed for a representative case in the mutual-friction-dominated regime ($T=2.0~\mathrm{K}$, $q=300~\mathrm{mW/cm^2}$, $B=30\%$). The resolution $\Delta x$ refers to the nominal cell size in the free-stream region, while finer resolutions are applied near boundaries.}
\begin{ruledtabular}
\begin{tabular}{cccl}
Grid & Nominal Resolution ($\Delta x=\Delta y$) & Total Cells & Predicted Topology \\
\hline
M1 & $2.0~\mathrm{mm}$ & 1,202 & 2-Vortex \\
M2 & $1.0~\mathrm{mm}$ & 4,384 & Disordered \\
M3 & $0.5~\mathrm{mm}$ & 16,896 & 4-Vortex \\
\textbf{M4} & \textbf{$0.25~\mathrm{mm}$} & \textbf{86,464} & \textbf{4-Vortex (converged)} \\
M5 & $0.125~\mathrm{mm}$ & 345,856 & 4-Vortex \\
\end{tabular}
\end{ruledtabular}
\end{table*}

\section{Numerical Implementation and Validation}
The coupled two-fluid and Vinen equations are solved using \textit{SFHeliumFOAM}, a custom solver developed within the open-source \textsc{OpenFOAM-8} environment~\cite{openfoam_weller}. Pressure--velocity coupling is handled using the PIMPLE algorithm, which combines PISO and SIMPLE to provide robust convergence for transient flows with strongly coupled and stiff source terms~\cite{Issa1986}. We employ second-order accurate discretizations to resolve the multi-scale wake structures and the sharp spatial variations of the mutual-friction field. Time integration uses the Crank--Nicolson scheme for the transient terms. In the momentum equations, the convective terms are discretized with the Gauss linearUpwind scheme to minimize numerical diffusion and preserve coherent wake features, while gradient and Laplacian operators are treated with the standard second-order Gauss linear central differencing scheme. Because the Vinen equation contains stiff nonlinear source terms, numerical stability is maintained through adaptive time stepping with a conservative Courant--Friedrichs--Lewy (CFL) constraint~\cite{Ferziger2002}; throughout the simulations we enforce $\mathrm{CFL}\le 0.1$ to prevent non-physical oscillations in $L$ and the resulting mutual-friction field.

The simulations are conducted in a two-dimensional rectangular domain that reproduces the geometry of the experimental channel~\cite{Chagovets2013,Zhang2005}. The domain length is fixed at $L_c=200~\mathrm{mm}$ and the width at $H=20~\mathrm{mm}$. The flow is aligned with the $y$-axis, and a circular cylinder of diameter $D$ is centered at $y=100~\mathrm{mm}$ and $x=0~\mathrm{mm}$. The cylinder diameter is varied from $2~\mathrm{mm}$ to $10~\mathrm{mm}$ to cover the experimental blockage ratios $B\equiv D/H=10\%$ to $50\%$. The extended upstream and downstream lengths ($100~\mathrm{mm}$ each) ensure that boundary effects do not interfere with the wake dynamics in the region of interest~\cite{Bertolaccini2017,Hu2025a}.

Boundary conditions are formulated to mimic the physical connection to the saturated He~II bath. At the inlet ($y=0$), a uniform normal-fluid velocity is prescribed, determined by the applied heat flux $q$ as $|\mathbf{v}_n|=q/(\rho s T)$. To enforce the zero net mass-flux condition of thermal counterflow, the superfluid velocity is set to oppose the normal flow with magnitude $|\mathbf{v}_s|=(\rho_n/\rho_s)|\mathbf{v}_n|$. At the outlet ($y=L$), the temperature is clamped to the bath temperature $T=T_{\mathrm{bath}}$, and the reference pressure is set to the saturated vapor pressure at $T_{\mathrm{bath}}$; for all other variables, a Neumann (zero-gradient) condition is applied. The channel walls ($x=\pm H/2$) and the cylinder surface are modeled as adiabatic boundaries.

Because the mutual-friction force varies sharply with the local counterflow, $F_{ns}\propto|\mathbf{v}_n-\mathbf{v}_s|^3$, adequate near-obstacle resolution is required, especially near the cylinder shoulders where the local velocity surge forms a narrow dissipative barrier. If the mesh is too coarse there, numerical diffusion can smear this localized force field and bias the resulting wake topology. We therefore use a multi-block structured mesh with an O-grid around the cylinder to maintain good grid orthogonality, with clustering near the cylinder surface and the channel side walls to resolve near-boundary gradients. A coarsened buffer zone is included near the outlet ($y>195~\mathrm{mm}$) to reduce spurious reflections from the boundary.

\begin{figure*}[t]
\includegraphics[width=1.0\linewidth]{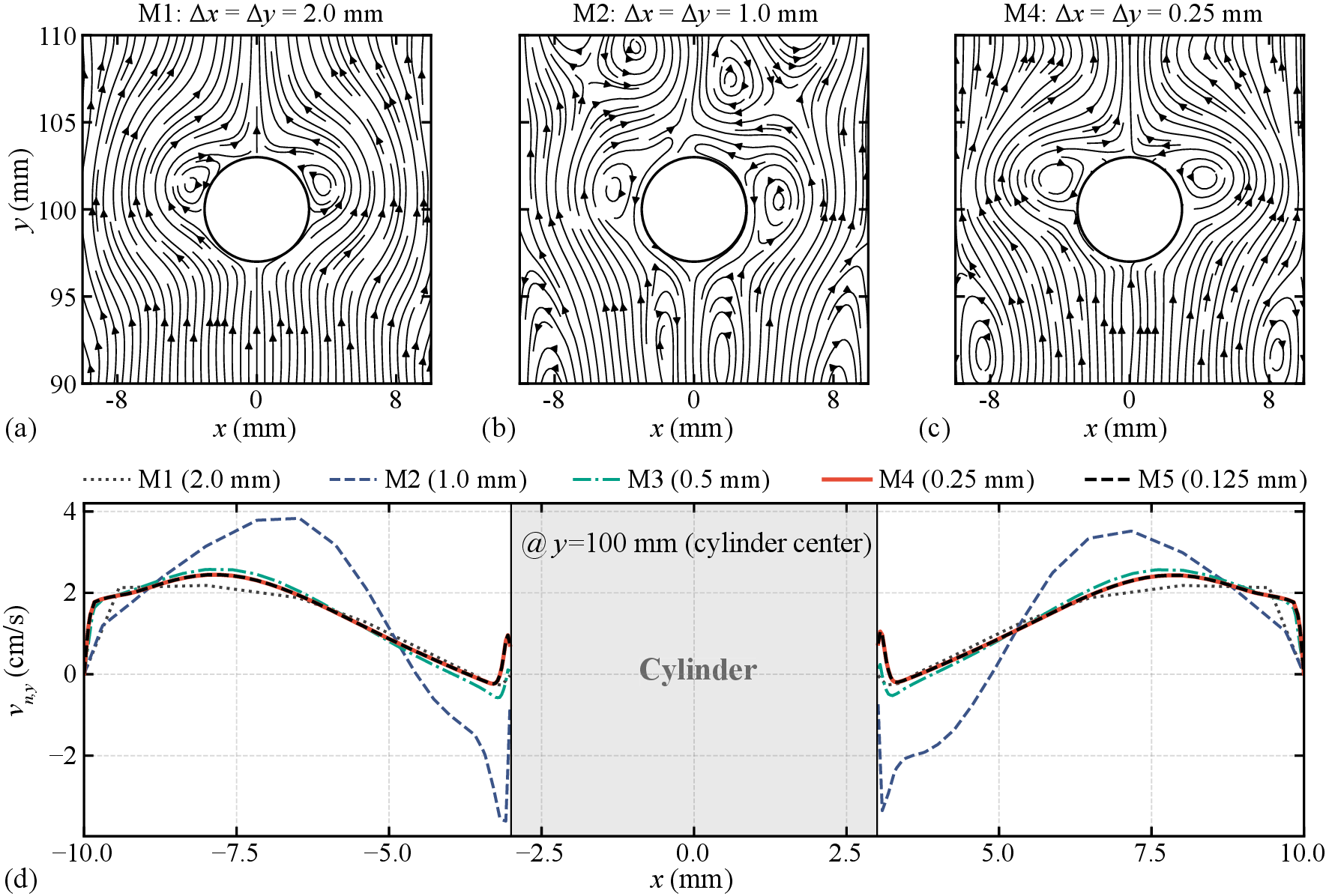}
\caption{\label{fig:S1} Grid-independence test for a representative case ($T=2.0~\mathrm{K}$, $q=300~\mathrm{mW/cm^2}$, $B=30\%$). (a)--(c) Comparison of streamline topologies obtained with different nominal grid resolutions. The experimentally observed 4-vortex state is robustly resolved for $\Delta x=\Delta y\le 0.5~\mathrm{mm}$. (d) Transverse profiles of the streamwise normal-fluid velocity $v_{n,y}$ along the channel width $x$ at the cylinder center ($y=100~\mathrm{mm}$). The profiles demonstrate clear convergence between the $0.25~\mathrm{mm}$ and $0.125~\mathrm{mm}$ grids.}
\end{figure*}

To determine the required resolution, we performed a systematic grid-independence study for a representative case in the mutual-friction-dominated regime ($T=2.0~\mathrm{K}$, $q=300~\mathrm{mW/cm^2}$, $B=30\%$), using five structured grids with nominal spatial resolutions ranging from $\Delta x=\Delta y=2.0~\mathrm{mm}$ to $0.125~\mathrm{mm}$ (Table~\ref{tab:grid}). As shown in Fig.~\ref{fig:S1}(a)--(c), the obtained topology depends strongly on grid size: at $\Delta x=\Delta y=2.0~\mathrm{mm}$ the solution relaxes to an erratic 2-vortex pattern, whereas at $\Delta x=\Delta y=1.0~\mathrm{mm}$ the flow becomes disordered. With further refinement, the anomalous 4-vortex state appears at $\Delta x=\Delta y=0.5~\mathrm{mm}$ and remains unchanged for $\Delta x=\Delta y\le 0.5~\mathrm{mm}$, becoming well defined and robust at $\Delta x=\Delta y=0.25~\mathrm{mm}$. Convergence is further quantified in Fig.~\ref{fig:S1}(d), where transverse profiles of the streamwise normal-fluid velocity $v_{n,y}(x)$ at the cylinder center ($y=100~\mathrm{mm}$) collapse between the $\Delta x=\Delta y=0.25~\mathrm{mm}$ and $0.125~\mathrm{mm}$ meshes. Based on these results, we adopt $\Delta x=\Delta y=0.25~\mathrm{mm}$ (Grid M4) for all production simulations reported in the main text.

\begin{figure}[t]
\includegraphics[width=1.0\linewidth]{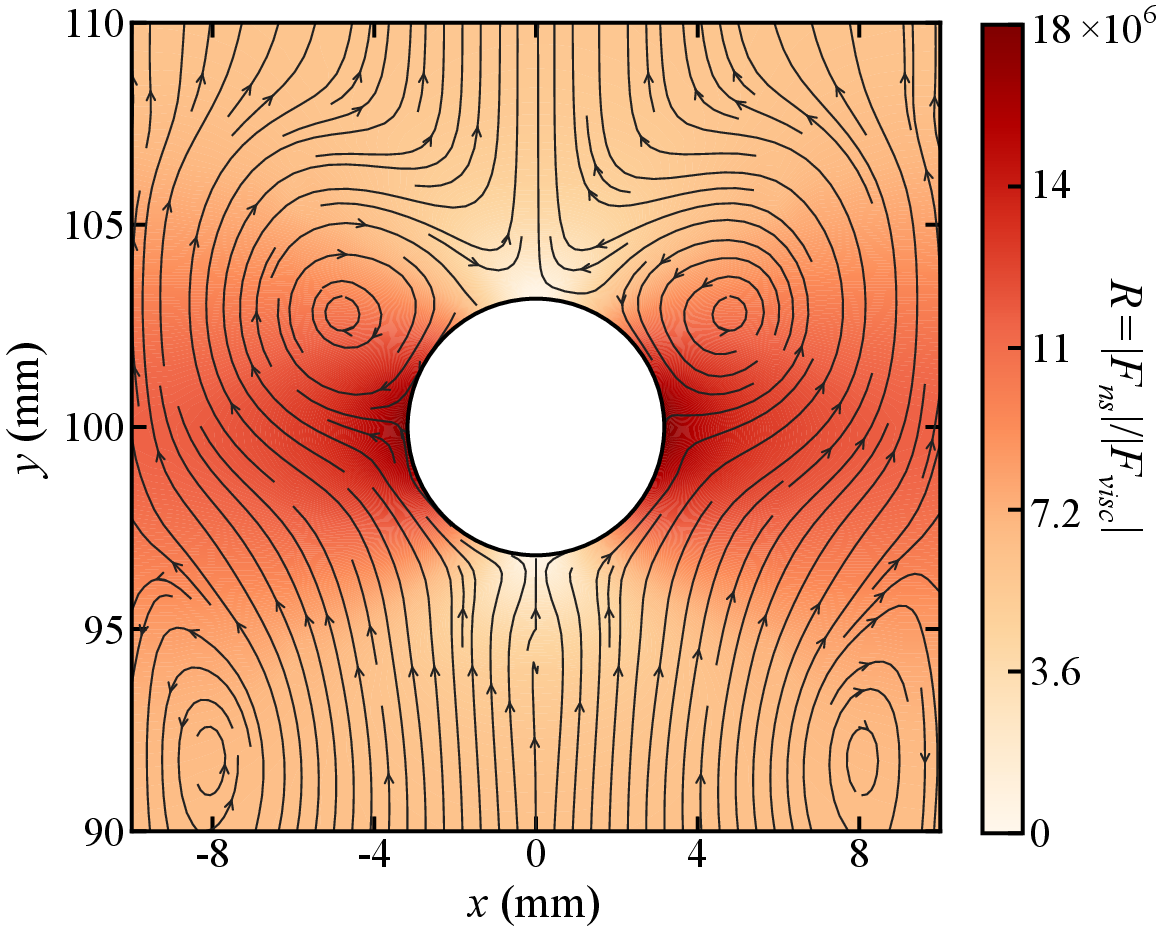}
\caption{\label{fig:S2} Contour map of the force ratio $R=|\mathbf{F}_{ns}|/|\mathbf{F}_{visc}|$ for the representative 4-vortex case in Fig.~1(d) of the main text ($T=2.03~\mathrm{K}$, $q=1120~\mathrm{mW/cm^2}$, $B=31.75\%$). Streamlines indicate the normal-fluid velocity field.}
\end{figure}

Finally, we verified that the 2-vortex state observed in the low-$q$ (inertia-dominated) regime is physically robust. Unlike the mutual-friction-dominated regime, where insufficient resolution can smear the localized mutual-friction barrier and alter the apparent topology, the low-$Re_n$ 2-vortex wake persists unchanged when computed on the finest grids (M4 and M5). This confirms that the $0$-vortex to $2$-vortex transition reported in the main text is a genuine physical result driven by inertia rather than a numerical artifact.

\section{Cylinder wake stabilization}
An intriguing feature of He~II counterflow wakes is their remarkable downstream stability: once the downstream vortices appear (and at higher heat fluxes), the wake remains steady rather than developing classical K\'arm\'an shedding. This behavior indicates that the wake instability is not set primarily by the balance between inertia and viscous diffusion. Instead, the vortex tangle introduces mutual friction as an additional dissipation channel, and in the relevant counterflow regime this vortex-mediated damping can overwhelm the normal-fluid viscous force in the regions that control separation and shedding.

To make this statement quantitative, we compare the mutual-friction force density $\mathbf{F}_{ns}=(\rho_n\rho_s/3\rho)\,B_L\,\kappa\,L\,\mathbf{v}_{ns}$ to the normal-fluid viscous force density $\mathbf{F}_{\mathrm{visc}}\equiv\nabla\cdot(\eta_n\nabla\mathbf{v}_n)$. A dimensionless ratio $R$ can be introduced:
\begin{equation}
R(\mathbf{x}) \equiv \frac{\left|\mathbf{F}_{ns}(\mathbf{x})\right|}{\left|\mathbf{F}_{\mathrm{visc}}(\mathbf{x})\right|}
= \frac{\left|\frac{\rho_n\rho_s}{3\rho}\,B_L\,\kappa\,L(\mathbf{x})\,\mathbf{v}_{ns}(\mathbf{x})\right|}
{\left|\nabla\cdot\big(\eta_n\nabla\mathbf{v}_n(\mathbf{x})\big)\right|},
\label{eq:R_ratio}
\end{equation}
which directly measures, point by point, whether dissipation is controlled by mutual friction or by viscosity.

We evaluate $R$ for the representative 4-vortex case shown in Fig.~1(d) of the main text ($T=2.03~\mathrm{K}$, $q=1120~\mathrm{mW/cm^2}$, $B=31.75\%$). As shown in Fig.~\ref{fig:S2}, $R$ is already $\gg 1$ across essentially the entire computational domain, demonstrating that vortex-mediated dissipation dominates viscous diffusion throughout this regime. Moreover, $R$ is nonuniform and exhibits a maximum near the cylinder shoulders, where the local counterflow intensifies and the vortex-line density is strongly amplified. This force hierarchy provides a direct explanation for the stable wake. In classical viscous wakes, shear-layer roll-up and K\'arm\'an shedding arise when inertial amplification outpaces viscous damping. Here, the separated shear layers experience strong mutual-friction damping instead, which suppresses the instability that would otherwise trigger periodic shedding.

\bibliography{reference}